# DISCUSSION OF: TREELETS—AN ADAPTIVE MULTI-SCALE BASIS FOR SPARSE UNORDERED DATA

By Fionn Murtagh

*University of London*

The work of Lee et al. is theoretically well founded and thoroughly motivated by practical data analysis. The algorithm presented has the following important properties:

1. Hierarchical clustering using a novel, adaptive, eigenvector-related, agglomerative criterion.
2. Principal components analysis carried out locally, leading to the required sample size for consistency being logarithmic rather than linear; and computational time being quadratic rather than cubic.
3. Multiresolution transform with interesting characteristics: data-adaptive at each node of the tree, orthonormal, and the tree decomposition itself is data-adaptive.
4. Integration of all of the following: hierarchical clustering, dimensionality reduction, and multiresolution transform.
5. Range of data patterns explored, in particular, block patterns in the covariances, and "model" or pattern contexts.

While I admire the work of the authors, nonetheless I have a different point of view on key aspects of this work:

1. The highest dimensionality analyzed seems to be 760 in the Internet advertisements case study. In fact, the quadratic computational time requirements (Section 2.1 of Lee et al.) preclude scalability. My approach in Murtagh (2007a) to wavelet transforming a dendrogram is of linear computational complexity (for both observations, and attributes) in the multiresolution transform. The hierarchical clustering, to begin with, is typically quadratic for the $n$ observations, and linear in the $p$ attributes. These computational requirements are necessary for the "small $n$, large $p$" problem which motivates this work (Section 1). In particular, linearity in $p$ is a *sine qua non* for very high dimensionality data exploration.

Since $L = O(p)$ in Section 2.1, this cubic time requirement has to be alleviated, in practice, through limiting $L$ to a user-specified value.

2. The local principal components analysis (Section 2.1) inherently helps with data normalization, but it only goes some distance. For qualitative, mixed quantitative and qualitative, or other forms of messy data, I would use a correspondence analysis to furnish a Euclidean data embedding. This, then, can be the basis for classification or discrimination, benefiting from the Euclidean framework. See Murtagh (2005).

3. My final point is in relation to the following (Section 1): "The key property that allows successful inference and prediction in high-dimensional settings is the notion of sparsity." I disagree, in that sparsity of course can be exploited, but what is far more rewarding is that high dimensions are of particular *topology*, and not just data *morphology*.

   This is shown in the work of Hall et al. (2005), Ahn et al. (2007), Donoho and Tanner (2005) and Breuel (2007), as well as Murtagh (2004). What this leads to, potentially, is the exploitation of the remarkable simplicity that is concomitant with very high dimensionality: Murtagh (2007b). Applications include text analysis, in many varied applications, and high frequency financial and other signal analysis.

In conclusion, I thank the authors for their thought-provoking and motivating work.

## REFERENCES

AHN, J., MARRON, J. S., MULLER, K. M. and CHI, Y.-Y. (2007). The high-dimension, low-sample-size geometric representation holds under mild conditions. *Biometrika* **94** 760–766.

BREUEL, T. M. (2007). A note on approximate nearest neighbor methods. Available at http://arxiv.org/pdf/cs/0703101.

DONOHO, D. L. and TANNER, J. (2005). Neighborliness of randomly-projected simplices in high dimensions. *Proc. Natl. Acad. Sci. USA* **102** 9452–9457. MR2168716

HALL, P., MARRON, J. S. and NEEMAN, A. (2005). Geometric representation of high dimension low sample size data. *J. Roy. Statist. Soc. B* **67** 427–444. MR2155347

MURTAGH, F. (2004). On ultrametricity, data coding, and computation. *J. Classification* **21** 167–184. MR2100389

MURTAGH, F. (2005). *Correspondence Analysis and Data Coding with R and Java*. Chapman and Hall/CRC, Boca Raton, FL. With a foreword by J.-P. Benzécri. MR2155971

MURTAGH, F. (2007a). The Haar wavelet transform of a dendrogram. *J. Classification* **24** 3–32. MR2370773

MURTAGH, F. (2007b). The remarkable simplicity of very high dimensional data: Application of model-based clustering. Available at www.cs.rhul.ac.uk/home/fionn/papers.

DEPARTMENT OF COMPUTER SCIENCE
ROYAL HOLLOWAY
UNIVERSITY OF LONDON
EGHAM, SURREY TW20 0EX
UNITED KINGDOM
E-MAIL: fmurtagh@acm.org